\documentclass[12pt]{article}
\usepackage{epsfig}
 \usepackage{times}
\begin{document}

\def\bib#1{[{\ref{#1}}]}
\title{\bf Cosmological dynamics in tomographic probability representation}
\author{{V.~I.~Man'ko${ }^{1,2}$, G.~Marmo${ }^{1,2}$ and C.~Stornaiolo ${ }^{1,2}$}\\
{\em $~^{1}$ Istituto Nazionale di Fisica Nucleare,}
{\em Sezione di Napoli,}\\
 {\em  Complesso Universitario di Monte S. Angelo}
 {\em Edificio N' }\\ {\em via Cinthia, 45 -- 80126 Napoli}\\
{\em $~^{2}$ Dipartimento di Scienze
Fisiche,}\\ {\em Universit\`{a} ``Federico II'' di Napoli,}\\
 {\em  Complesso Universitario di Monte S. Angelo}
 {\em Edificio N'  }\\ {\em via Cinthia, 45 -- 80126  Napoli}\\ }
\date{ }
\maketitle

\begin{abstract}
The probability representation for quantum states of the universe
in which the states are described by a fair probability
distribution instead of wave function (or density matrix) is
developed to consider cosmological dynamics. The evolution of the
universe state is described by standard positive transition
probability (tomographic transition probability) instead of the
complex transition probability amplitude (Feynman path integral)
of the standard approach. The latter one is expressed in terms of
the tomographic transition probability. Examples of
minisuperspaces in the framework of the suggested approach are
presented. Possibility of observational applications of the
universe tomographs are discussed.
\end{abstract}

\vskip 0.5truecm
\section{Introduction}

Recently \bib{Manko:2003dp} tomographic probability approach to
describe the states of the universe in quantum cosmology was
suggested. In the framework of this approach the quantum state of
the universe is associated with the standard positive probability
distribution (function or functional). The probability
distribution contains the same information on the universe quantum
state that  the wave function of the universe \bib{hawking1}
\bib{whdw}\bib{hh} or the density matrix of the universe  \bib{page},
\bib{hawking2}. The
latter can be presented in different forms, e.g. in form of a
Wigner function \bib{wigner32}  considered in  \bib{anton} in a
cosmological context. In fact the tomographic probability
distribution describing the state of the universe is a symbol of a
density operator
\bib{marmops}\bib{marmoolga} and the tomographic symbols of the
operators realize one of the variants of the star-product
quantization scheme widely used \bib{fronsdal} to study the
relation of classical and quantum pictures \bib{marmo2}, which can
also be applied to study the relation of classical and quantum
descriptions of the universe in quantum cosmology. One of the
important ingredients of such descriptions is the evolution of the
state. In quantum mechanics such evolution is completely described
by means of a complex transition probability amplitude from an
initial state to a final one. This probability amplitude
(propagator) can be presented in the form of a Feynman path
integral containing the classical action. In quantum mechanics in
the probability representation using the tomographic approach the
state evolution can be associated with the standard transition
probability. It contains also information on the transition
probability amplitude related to the probability by integral
transform induced by the Radon transform relating the density
matrix  (Wigner function) with the quantum tomographic probability
\bib{mancini}, \bib{lecture},  \bib{mendespl}, \bib{mendes2}.

In our previous work
\bib{Manko:2003dp} we suggested to associate the state of the
universe in quantum cosmology with the tomographic probability (or
tomogram). The aim of our paper is to consider now in the
framework of the suggested probability representation of the
universe state in quantum cosmology also the cosmological dynamics
and to express this dynamics in terms of a positive transition
probability connecting initial and final tomograms of the
universe. Another goal of the work is to discuss a possible
experimental approach to observe the tomogram of the universe  at
its present stage and try to extract some information on the
tomogram of the initial state of the universe. The idea of this
attempt is based on the fact that the same tomogram describes the
state of a classical system and its quantum counterpart. In this
sense in the probability representation of the quantum state there
is no such  dramatic difference between the classical and quantum
pictures as the difference between wave function (or density
matrix) and classical probability distribution (or trajectory) in
the classical phase space. Due to this one can try to study the
cosmological dynamics namely in the tomographic probability
representation.

In order to illustrate the idea we will use the same simple
example of the universe description by means of the minisuperspace
discussed, e.g. in  \bib{hh}, \bib{Lemos:1995qu} . In these
minisuperspaces the quantum cosmological dynamics in operative
form is reduced to the dynamics of formal quantum systems
described by Hamiltonians of the types of oscillator, free motion
and free falling particles. In view of this one can apply the same
recent obtained results on description of  such systems by
tomographic probabilities  \bib{olga} to the cosmological
dynamics.

The paper is organized as follows. In the next section we will
review the cosmology in terms of a homogeneous (and isotropic)
metric with a time dependent parameter the expansion factor of the
universe. In section 3, we review the tomographic approach to
evolution of the quantum system. In section 4 we consider the
examples of the minisuperspace described by the reduced
Hamiltonians.

\section{Cosmology}

Classically a homogeneous and isotropic universe is described by
one of the following  metrics
\begin{equation}\label{metric}
    ds^{2}=-c^{2}dt^{2}+
    \frac{ a^{2}}{1-kr^{2}}\left(
    dr^{2}+r^{2}d\theta^{2}+r^{2}\sin^{2}\theta\right)d\phi^{2}
\end{equation}
where the parameter $k$ can be $k>0$, $k=0$ and $k<0$ being
related to a closed universe, a flat universe and an open universe
respectively.

When the gravitational source is a perfect fluid, described by the
energy-momentum tensor, the Einstein equations with the metric
(\ref{metric}) are in the second order form

\begin{equation}\label{einstein1}
  \frac{\ddot{a}}{a}=-\frac{4\pi G}{3}(\rho+3P)
\end{equation}
which represents the dynamic equation and

\begin{equation}\label{einstein2}
    \frac{{\dot{a} }^{2}}{{a}^{2}}+\frac{k}{a^{2}}=\frac{8\pi G}{3}\rho
\end{equation}
which is a constraint, i.e. it defines the manifold of allowed
initial conditions. It takes a simple computation to show that
there are no secondary constraints. It constitutes an ``invariant
relation'', according to Levi-Civita.

From equations (\ref{einstein1}) and (\ref{einstein2}) the first
order equation
\begin{equation}\label{conservation}
     \dot{\rho}=-3\frac{\dot{a}}{a}\left(\rho+P\right)
\end{equation}
can be derived by taking the time derivative of (\ref{einstein2}).
It can be used alternatively in a system with equation
(\ref{einstein2}).

The system of equations (\ref{einstein1}) and (\ref{einstein2}) or
   (\ref{einstein2}) and (\ref{conservation}) are not complete, they  must be
completed by an equation of state $P=P(\rho)$ which is generally
linear, $P= (\gamma-1)\rho$ ($\gamma=1$ is the so-called matter
fluid, $\gamma=4/3$ is the radiation fluid and so on).

Equation (\ref{conservation}) together with an equation of state,
is important for our purpose  because it shows that the lefthand
side of equations (\ref{einstein1}) and (\ref{einstein2}) can be
expressed as a function of $a$ and represents a force in these
equations, if we treat them as equations for a ``point'' particle
as a result we have

 \begin{equation}\label{rhodia}
    \rho= \frac{\rho_{0}a_{0}^{3\gamma}}{a^{3\gamma}}
\end{equation}
when the equation of state is linear.

It is possible to derive the cosmological model from a point
particle Lagrangian, where the expansion factor $a$ takes the part
of the particle coordinate. Let us introduce the following
Lagrangian

\begin{equation}\label{lagrangian}
   \mathcal{L}=3a\dot{a}^{2}-3ka - 8\pi G \rho_{0}a_{0}^{3\gamma}a^{3(1-\gamma)}.
\end{equation}

The gravitational part is formally derived by substituting
directly metric (\ref{metric}) into the (field) general
relativistic action $\int\sqrt{-g}R$ and the material part is
obtained by putting a corresponding potential term $\Phi(a)=8\pi G
\rho_{0}a_{0}^{3\gamma}a^{3(1-\gamma)}$, in the case of a fluid
source.

 Equation (\ref{einstein1}) follows from the variation of the
Lagrangian (\ref{lagrangian}).

From equation (\ref{lagrangian}) the conjugate momentum of $a$ is
\begin{equation}\label{momentum}
     p_{a}=\frac{\partial L}{\partial\dot{ a}}= 6 a \dot{a}.
\end{equation}

 Equation (\ref{einstein2}) is a constraint which is equivalent to the vanishing of the ``energy
 function'' $ E_{\mathcal{L}}$ associated to the Lagrangian
 \begin{equation}\label{energyfunction}
    E_{\mathcal{L}}=3a\dot{a}^{2}+ 3ka -8\pi G \rho_{0}a_{0}^{3\gamma}a^{3(1-\gamma)}.
\end{equation}

An alternative way to describe cosmology with a cosmological
fluid, with $\Lambda=0$,  was introduced by Lemos
\bib{Lemos:1995qu} and Faraoni \bib{Faraoni:1999qu}

They showed that equations (\ref{einstein1})  and
(\ref{einstein2}) can be transformed in equations similar to the
harmonic oscillator ones. By passing to the conformal time $\eta$,
defined by the relation
$$d\eta=\frac{dt}{a(t)},$$ and with the change of variables
 \begin{equation}\label{substitution}
w= {a}^{\chi}
\end{equation} where $$\chi=\frac{3}{2}\gamma-1 $$
equation (\ref{einstein1}) takes the form
\begin{equation}\label{harmonic}
    w''+k\chi^{2} w=0.
\end{equation}

Let us consider now a cosmological model where the source is
originated by a scalar field, which satisfies the Klein-Gordon
equation specialized to a homogeneous and isotropic universe
(which substitutes equation (\ref{conservation}))

\begin{equation}\label{kleingordon}
    \ddot{\varphi}+3\frac{\dot{a}}{a}\dot{\varphi} + V'(\varphi)=0
\end{equation}
where $V(\varphi)$ is the potential for the scalar field.

It is possible to derive the cosmological equations
(\ref{einstein1}) and (\ref{einstein2}), where $\rho_{\varphi}=1/2
\dot{\varphi}^{2}+V(\varphi)$ and $P_{\varphi}=1/2
\dot{\varphi}^{2}-V(\varphi)$  from a Lagrangian
$\mathcal{L}=\mathcal{L}(a,\dot{a}, \varphi,\dot{\varphi})$.
\begin{equation}\label{lagrangian1}
   \mathcal{L}=3a\dot{a}^{2}-3ka -8\pi
   G a^{3}\left(\frac{1}{2}\dot{\varphi}^{2}- V(\varphi)\right).
\end{equation}

where the minisuperspace becomes a two dimensional space with
coordinates $a$ and $\varphi$.

From the previous Lagrangian (\ref{lagrangian1}) we compute the
conjugate momenta  of $a$ and $\varphi$:
\begin{equation}\label{momentuma}
   p_{a}=  \frac{\partial L}{\partial \dot{a}}= 6 a \dot{a}.
\end{equation}
\begin{equation}\label{momentumfi}
 p_{\varphi}=\frac{\partial L}{\partial \dot{\varphi}}=8\pi
   G a^{3}\dot{\varphi}
\end{equation}
 Equation (\ref{einstein2}) is now a constraint which is equivalent to the vanishing of the ``energy
 function'' $ E_{\mathcal{L}}$ associated to the Lagrangian
 \begin{equation}\label{energyfunction}
    E_{\mathcal{L}}=3a\dot{a}^{2}+ 3ka -8\pi
   G a^{3}\left(\frac{1}{2}\dot{\varphi}^{2}+ V(\varphi)\right).
\end{equation}

If  $V(\varphi)$ takes the form
\begin{equation}\label{potperfect}
    V(\varphi)=\frac{1}{2}(2-\gamma)\rho^{0}_{\varphi}\exp(-\sqrt{24\pi
    G\gamma}\varphi),
\end{equation}
with constant $\gamma$, then $\rho_{\varphi}$ and $P_{\varphi}$
satisfy the equation of state \bib{Stornaiolo:1994mw}
\begin{equation}\label{eos}
 P_{\varphi}= (\gamma-1)\rho_{\varphi}
\end{equation}
and also in this case, the evolution of universe can be described
by  equation (\ref{harmonic}). In \bib{gousheh} there are other
examples in which the cosmological models with a scalar field can
be described by equations similar with (\ref{harmonic}), the
tomographic version of these models has already been discussed in
\bib{Manko:2003dp}.

\section{Evolution in minisuperspace in the framework of tomographic probability representation}

We will discuss below the evolution of a universe in the framework
of the minisuperspace model discussed in the previous section.
Thus the state of the universe is described by a wave function
$\Psi(x,t)$. This wave function evolves in time from its initial
value $\Psi(x,t_{0})$ and this evolution can be described by a
propagator $G(x,x',t,t_{0})$
\begin{equation}\label{propagatorevol}
\Psi(x,t)=\int G(x,x',t,t_{0})\Psi(x',t_{0})dx'.
\end{equation}
The propagator can be obtained using path integration over
classical trajectories of the exponential of the classical action
$S$
\begin{equation}\label{propagatoraction}
G(x,x',t,t_{0})= \int D[x(t)] e^{\frac{iS[x(t)]}{\hbar}}.
\end{equation}

In our previous work \bib{Manko:2003dp} we discussed the
properties of the new representation (tomographic probability
representation) of the quantum states of the universe.

In this representation (which we discuss below in the framework of
a minisuperspace model) the wave function of the universe
$\Psi(x,t)$ or the density matrix of the universe

\begin{equation}\label{rho}
\rho(x,x',t)=\Psi(x,t)\Psi^{*}(x',t)
\end{equation}

can be mapped onto the standard positive distribution ${\cal W}(X,
\mu,\nu, t)$ of the random variable $X$ depending on the two real
extra parameters $\mu$ and $\nu$ and the time $t$. The map is
given by the formula (we take $\hbar=1)$
\begin{equation}\label{probdistribution}
    {\cal W}(X, \mu,\nu, t)= \frac{1}{2\pi |\nu|}\int\rho(y,y',t)
    e^{i\frac{\mu(y^{2}-{y'}^{2})}{2\nu}-i\frac{X}{\nu}(y-y')}dy
    dy'.
\end{equation}

In fact, equation (\ref{probdistribution}) is the fractional
Fourier transform \bib{margarita} \bib{marg} of the density
matrix. The map has inverse and the density matrix can be
expressed in terms of the tomographic probability representation
as follows

\begin{equation}\label{densitytomograph}
    \rho(x,x',t)=\frac{1}{2\pi} \int
    \mathcal{W}(Y,\mu,x-x',t)\,
    e^{i\left(Y-\frac{\mu}{\nu}(x+x')\right)}dYd\mu.
\end{equation}
The expression (\ref{probdistribution}) can be given in an
invariant form \bib{mendes3}
\begin{equation}\label{invariantform}
  {\cal W}(X, \mu,\nu, t)= \left\langle \delta(X-\mu\hat{q}-\nu\hat{p})\right\rangle
\end{equation}
Here $\langle\ \rangle$ means trace with the density operator
$\hat{\rho} (t)$ of the universe state, $\hat{q}$ and $\hat{p}$
are the operators of position (universe expansion factor) and the
conjugate moment respectively. From equation (\ref{invariantform})
some properties of the tomogram ${\cal W}(X, \mu,\nu, t)$ are
easily extracted. First, the universe tomogram is a normalized
probability distribution, i.e.
\begin{equation}\label{normalized}
    \int{\cal W}(X,
\mu,\nu, t) dX=1
\end{equation}
if the universe density operator is normalized (i.e. $Tr
\hat{\rho}(t)=1$). Second, the tomogram of the universe state has
the homogeneity property \bib{rosapl}
\begin{equation}\label{homogeneity}
    {\cal W}(\lambda X,
\lambda\mu,\lambda\nu, t)= \frac{1}{|\lambda|} {\cal W}(X,
\mu,\nu, t)
\end{equation}
The tomogram can be related with  such quasidistribution as the
Wigner function $W(q,p,t)$ \bib{wigner32} used in the phase space
representation of the universe states in \bib{anton}.

The relation reads

\begin{equation}\label{wignertomo}
  {\cal W}(X,\mu,\nu,t)= \int W(q,p,t)\delta(X-\mu q - \nu p)\frac{dqdp}{2\pi}
\end{equation}

which is the standard Radon transform of the Wigner function. The
physical meaning of the tomogram $ {\cal W}(X,\mu,\nu,t)$ is the
following. One has in the phase space the line
\begin{equation}\label{line}
    X=\mu q + \nu p
\end{equation}
which is given by the zero of the delta-function argument in
equation (\ref{wignertomo}). The real parameters $\mu$ and $\nu$
can be given in the form
\begin{equation}\label{parameters}
    \mu=s\cos\theta~~~~~~~~~~~~~~~~~\nu=s^{-1}\sin\theta.
\end{equation}
Here $s$ is a real squeezing parameter and $\theta$ is a rotation
angle. Then the variable $X$ is identical to the position measured
in the new reference frame in the universe phase-space. The new
reference frame has new scaled axis $sq$ and $s^{-1}p$ and after
the scaling the axis are rotated by an angle $\theta$. Thus the
tomogram implies the probability distribution of the random
position $X$ measured in the new (scaled and rotated) reference
frame in the phase-space. The remarkable property of the
tomographic probability distribution is that being a fair positive
probability distribution, it contains a complete information of
the universe state contained in  the  density operator
$\hat{\rho}(t)$ which can be expressed in terms of the tomogram as
\bib{d'ariano}
\begin{equation}\label{operatorrho}
    \hat{\rho}(t)=\frac{1}{2\pi}\int \mathcal{W}(X,\mu,\nu,t)e^{i(X-\mu\hat{q}-\nu\hat{p})}
dX d\mu d\nu
\end{equation}
Formulae (\ref{invariantform}) and (\ref{operatorrho}) can be
treated with the tomographic star-product quantization schemes
\bib{marmoolga} used to map the universe quantum observables (operators) onto
functions (tomographic symbols) on a manifold $(X,\mu,\nu)$. The
tomographic map can be used not only for the description of the
universe state by probability distributions, but also to describe
the evolution of the universe (quantum transitions) by means of
the standard real positive transition probabilities (alternative
to the complex transition probability amplitudes). The transition
probability
$$\Pi(X,\mu,\nu,t,X',\mu',\nu',t_{0})$$ is the propagator which
gives the tomogram of the universe  ${\cal W}(X,\mu,\nu,t)$, if
the tomogram at the initial time $t_{0}$ is known, in the form
\begin{equation}\label{proptomo}
     {\cal W}(X,\mu,\nu,t)= \int \Pi(X,\mu,\nu,t,X',\mu',\nu',t_{0}){\cal
     W}(X',\mu',\nu',t_{0})dX' d\mu'd\nu'.
\end{equation}
The positive transition probability describing the evolution of
the universe has the obvious nonlinear properties used in
classical probability theory, namely
$$ \Pi(X_{3},\mu_{3},\nu_{3},t_{3},X_{1},\mu_{1},\nu_{1},t_{1})=\int
 \Pi(X_{3},\mu_{3},\nu_{3},t_{3},X_{2},\mu_{2},
\nu_{2},t_{2})$$
\begin{equation}\label{postransprob}
~~~~~~~~~~~~~~~\times\Pi(X_{2},\mu_{2},\nu_{2},t_{2},X_{1},\mu_{1},\nu_{1},t_{1})\,dX_{2}\,d\mu_{2}\,d\nu_{2}.
\end{equation}
They follow from the associativity property of the evolution maps.
This nonlinear relation is analogous to the nonlinear relation of
the complex quantum propagators of the universe wave function
 \begin{equation}\label{complexnonlinear}
 G(x_{3},x_{1},t_{3},t_{1})=\int G(x_{3},x_{2},t_{3},t_{2})
 G(x_{2},x_{1},t_{2},t_{1})dx_{2}.
\end{equation}
Both relations (\ref{postransprob}) and (\ref{complexnonlinear})
imply that the state of the universe evolves from the initial one
  to the final one   through all
intermediate states. The remarkable fact is that this quantum
evolution of the universe state can be associated with the
standard positive transition probabilities like in classical
dynamics. This is connected with the existence of the invertible
relations of the tomographic and quantum propagators
\bib{mendes2}\bib{olga} . If one denotes
\begin{equation}\label{kappa}
    K(X,X',Y,Y',t)=G(X,Y,t)G^{*}(X',Y',t),
\end{equation}
then the quantum propagator may be given in the following form
$$ K(X,X',Y,Y',t)= \frac{1}{(2\pi)^{2}}\int \frac{1}{|Y'|}
\exp\left\{i\left(Y-\mu \frac{(X+X')}{2}
\right)-i\frac{Z-Z'}{\nu'}Y'\right.$$

\begin{equation}\label{kappa1}
\left. + i
\frac{Z^{2}-{Z'}^{2}}{2\nu'}\mu'\right\}\Pi(Y,\mu,X-X',0,X',\mu',\nu',t)d\mu\,d\mu'\,dY\,dY'\,d\nu'.
\end{equation}

This relation can be reversed. Thus the propagator for the
tomographic probability can be expressed in terms of the Green
function $G(x,y,t)$ as follows (we take $t_{0}=0$)
$$\Pi(X,\mu,\nu,X',\mu',\nu',t)=\frac{1}{4\pi}\int
    k^{2}G(a+\frac{k\nu}{2}, y,t)G^{*}(a-\frac{k\nu}{2},
    z,t)\delta(y-z-k\nu') $$

\begin{equation}\label{inverse}
   \times \exp\left[ik(X'-X+\mu q-\mu'\frac{y+z}{2})\right]dk dy dz
   dq.
\end{equation}
The relation can be used to express the tomographic propagation in
terms of the Feynmann path integral using the formula for the
quantum propagator (\ref{propagatoraction}) where the classical
action is involved. It means that the positive transition
probabilities (\ref{inverse}) can be reexpressed in terms of the
double path integral (with four extra usual integrations).

The discussed relations demonstrate that the  quantum universe
evolution can be described completely using only positive
transition probabilities.

Standard complex transition probability amplitudes (and Feynman
path integral) can be reconstructed using this transition
probability by means of equation (\ref{kappa1}.)

\section{Evolution of the universe in the oscillator model
framework}

As we have shown the equation for the universe evolution in the
conformal time picture (\ref{harmonic}) can be cast in  the form
of an oscillator equation. The oscillator has the frequency
$\omega^{2}=\pm k\chi^{2}$.

For $k=0$ one has the model of free motion. For $k<0$ one has the
model of a inverted oscillator and for $k>0$ one has the
standard oscillator as solution of the equation (\ref{harmonic}).
Since the problem of gravity quantization is not established with
complete rigor, we assume below that the quantum behavior of the
universe in the framework of the considered minisuperspace model
is described by the quantum behavior of the oscillator. Though the
connection (\ref{substitution}) of the expansion factor $a(\eta)$
with the classical observable $w$ which obeys to oscillator motion
provides constraints on the   ranging domain of this variable, we
assume in the quantum picture of the variable to lie on the real
line $R$. In such approach we apply the tomographic probability
representation, developed in the last section, to quantum states
of the universe in the framework of the oscillator model. We will
denote  in the quantum description the variable as $q$
 ($w\rightarrow q$)and the conformal time as $t$\  ($\eta\rightarrow
t$). Thus the tomographic probability $\mathcal{W}(X,\mu,\nu,t)$of
the universe state obeys the evolution equation
\bib{Manko:2003dp} for the potential energy $V(q)$ in the form
$$\frac{\partial\mathcal{W}(X,\mu,\nu,t)}{\partial t}-
     \mu\frac{\partial\mathcal{W}(X,\mu,\nu,t)}{\partial \nu}+
     i\left[V\left(-\left(\frac{\partial}{\partial X}\right)^{-1}\frac{\partial}{\partial\mu}
     -\frac{i\nu}{2}\frac{\partial}{\partial X}\right)\right.$$
\begin{equation}\label{tomoequation1}
\left. - V\left(-\left(\frac{\partial}{\partial
X}\right)^{-1}\frac{\partial}{\partial\mu}
     +\frac{i\nu}{2}\frac{\partial}{\partial
     X}\right)\right]\mathcal{W}(X,\mu,\nu,t)=0,
\end{equation}
where the operator $(\partial/\partial X)^{-1} $ is defined by the
relation
\begin{equation}\label{relation}
\left(\frac{\partial}{\partial X}\right)^{-1}\int
f(y)e^{iyX}dy=\int \frac{f(y)}{(iy)}e^{iyX}dy.
\end{equation}
The propagator of this equation $\Pi(X,\mu,\nu,t,X',\mu',\nu')$
satisfies equation (\ref{tomoequation1}) with the extra term
$$\frac{\partial\Pi}{\partial t}-
     \mu\frac{\partial\Pi}{\partial \nu}+
     i\left[V\left(-\left(\frac{\partial}{\partial X}\right)^{-1}\frac{\partial}{\partial\mu}
     -\frac{i\nu}{2}\frac{\partial}{\partial X}\right)\right.$$
\begin{equation}\label{tomoequation2}
\left. - V\left(-\left(\frac{\partial}{\partial
X}\right)^{-1}\frac{\partial}{\partial\mu}
     +\frac{i\nu}{2}\frac{\partial}{\partial
     X}\right)\right]\Pi=\delta(\mu-\mu')\delta(\nu-\nu')\delta(X-X')\delta(t),
\end{equation}
For the considered model the general equation for the universe
tomogram evolution takes the simple form of a first order
differential equation
\begin{equation}\label{tomoequation3}
\frac{\partial\mathcal{W}}{\partial t}-
     \mu\frac{\partial\mathcal{W} }{\partial \nu}
      + \omega^{2}\nu\frac{\partial\mathcal{W} }{\partial \mu} =0.
\end{equation}
Analogously for the propagator of the tomographic equation for the
universe in the framework of the oscillator model one has
\begin{equation}\label{tomoequation4}
\frac{\partial\Pi }{\partial t}-
     \mu\frac{\partial\Pi }{\partial \nu}
      + \omega^{2}\nu\frac{\partial\Pi }{\partial \mu} =\delta(\mu-\mu')\delta(\nu-\nu')\delta(X-X')\delta(t).
\end{equation}

The solution to this equation can be found to be in the case $k>0$
$$ \Pi^{osc.}(X,\mu,\nu,t,X',\mu',\nu')= \delta(X-X') \delta(\mu'-\mu\cos\,
\omega t+ \omega\nu\sin \omega t)$$
\begin{equation}\label{solutionk1}
\times \delta\left(\nu'-\nu\cos\, \omega t -
\frac{\mu}{\omega}\sin\, \omega t\right).
\end{equation}

In the limit $k=0$ (free motion) the equation for the tomogram
(\ref{tomoequation3}) becomes
\begin{equation}\label{tomoequation5}
\frac{\partial\mathcal{W}(X,\mu,\nu,t)}{\partial t}-
     \mu\frac{\partial\mathcal{W}(X,\mu,\nu,t)}{\partial \nu} =0.
\end{equation}

The corresponding propagator solution  reads

\begin{equation}\label{solutionk2}
  \Pi^{free}(X,\mu,\nu,t,X',\mu',\nu')= \delta(X-X')
\delta(\mu'-\mu) \delta(\nu'-\nu -\mu t).
\end{equation}
Finally for the case $k<0$   the propagator has the form
corresponding to a repulsive oscillator
$$ \Pi^{rep.}(X,\mu,\nu,t,X',\mu',\nu')= \delta(X-X') \delta(\mu'-\mu\cosh\,
\omega t -  \omega\nu\sinh \omega t)$$
\begin{equation}\label{solutionk3}
\times \delta\left(\nu'-\nu\cosh\, \omega t -
\frac{\mu}{\omega}\sinh\, \omega t\right).
\end{equation}
Thus we got the dynamics of the universe given by the transition
probabilities $\Pi^{osc.}$, $\Pi^{free}$ and $\Pi^{rep.}$ for the
three cases $k>0$, $k=0$ and $k<0$ respectively. One can see that
this dynamics is compatible with the dynamics calculated in the
standard representation of the complex Green function (quantum
propagator). For  $k=1$ the form of the Green function reads
\begin{equation}\label{Greenk1}
     G^{osc.}(X,X',t)=\sqrt{\frac{\omega}{2\pi i \sin \omega t}}
     \exp\left\{\frac{i\omega}{2}\left[\cot \omega t
     \left(X^{2}+{X'}^{2}\right)-\frac{2XX'}{\sin \omega t}\right]\right\}
\end{equation}
For the case of the free motion model the Green function can be
obtained by the limit $\omega\to 0$ in this expression and one has
\begin{equation}\label{Greenk2}
     G^{free}(X,X',t)=\sqrt{\frac{1}{2\pi i t}}
     \exp\left[i
     \frac{\left(X-{X'}\right)^{2}}{2t}\right]
\end{equation}
and for the repulsive oscillator model one has
\begin{equation}\label{Greenk3}
     G^{rep.}(X,X',t)=\sqrt{\frac{\omega}{2\pi i \sinh \omega t}}
     \exp\left\{\frac{i\omega}{2}\left[\coth \omega t
     \left(X^{2}+{X'}^{2}\right)-\frac{2XX'}{\sinh \omega
     t}\right]\right\}.
\end{equation}
All these three universe cases can be discussed using the Green
function in terms of the Feynmann path integral.

Thus the expression (\ref{Greenk1}) is given  by the formula
\begin{equation}\label{feynmanngreen}
     G(X,X't)=\int e^{i\int_{0}^{t}\left[\frac{\dot{x}^{2}(t)}{2}-
     \frac{\omega^{2}x^{2}(t)}{2}\right]dt}D[x(t)]
\end{equation}

The integral in the exponent of the path integral provides the
classical action for the oscillator
\begin{equation}\label{classical}
  S^{cl.}(X,X',t)= \int_{0}^{t}\left[\frac{\dot{x}^{2}(t)}{2}-
     \frac{\omega^{2}x^{2}(t)}{2}\right]dt
\end{equation}
where the trajectories start at $t=0$ at $X'$ and end at time $t$
at the point $X$. The classical action satisfies the
Hamilton-Jacobi equation

\begin{equation}\label{hamilton}
    \frac{\partial S^{cl.}(q,q',t)}{\partial t} +
    \mathcal{H}\left(q,p=- \frac{\partial S^{cl.}(q,q',t)}{\partial q}\right)=0
\end{equation}
where $\mathcal{H}$ is the Hamiltonian
\begin{equation}\label{hamiltonian}
 \mathcal{H}=\frac{p^{2}}{2}+\frac{\omega^{2}q^{2} }{2}
\end{equation}
For the free motion  model one has

\begin{equation}\label{freen}
  G^{free}(X,X',t)=\int
  e^{i\int_{0}^{t}\frac{\dot{x}^{2}(t)}{2}dt}D[x(t)].
\end{equation}
The path integral is integrated and the result (\ref{Greenk2})
contains in the exponent term the classical action
\begin{equation}\label{action}
     S^{(f)}(X,X',t)=\frac{(X-X')^{2}}{2t}
\end{equation}
which is solution of the Hamilton-Jacobi equation with the
Hamiltonian
\begin{equation}\label{freehamiltonian}
     \mathcal{H}=\frac{p^{2}}{2}.
\end{equation}
For the repulsive model one has the same structure of path
integral and the result of path integration is expressed in terms
of the classical action
\begin{equation}\label{classicalrepulsive}
    S^{rep.}(X,X',t)= \frac{ \omega}{2}\left[\coth \omega t
     \left(X^{2}+{X'}^{2}\right)-\frac{2XX'}{\sinh \omega
     t}\right],
\end{equation}
which is solution of the Hamilton-Jacobi equation with the
Hamiltonian
\begin{equation}\label{hrep}
    \mathcal{H}^{rep.}=\frac{p^{2}}{2}-\frac{\omega^{2}q^{2} }{2}.
\end{equation}
It is remarkable that all the obtained propagators complex Green
functions or path integrals are related with the propagators in
probability representation by means of equations
(\ref{postransprob}) and (\ref{complexnonlinear}).

Thus the universe evolution can be described in the oscillator
model of minisuperspace  for $k>0$, $k=0$ and $k<0$ by means of
the standard transition probabilities expressed as propagators
$\Pi^{osc.}$, $\Pi^{free}$ and $\Pi^{rep.}$ respectively.

\section{Conclusions}

To conclude we discuss the main results of the work. In addition
to what suggested in \bib{Manko:2003dp}, the probability
representation of the universe quantum states for which the states
(e.g. of the universe in a minisuperspace model) are described by
the standard positive probability distribution, we introduce the
description of the universe dynamics by means of standard
transition probabilities.

The transition probabilities are determined as propagators
(integral kernels) providing the evolution of the universe
tomograms. It is shown that there is a relation of the standard
propagator determining the quantum evolution of the universe wave
function to the tomographic propagator. This relation permits to
reconstruct the complex propagator  for the wave function in terms
of the positive propagator for the universe tomogram. Also, one
can express the propagator for the tomogram in terms of the
propagator for the wave function of the universe.

These relations between the propagators mean that the Feynmann
path integral formulation or the universe properties (in quantum
gravity) contains the same information that the probability
representation of the quantum states of the universe including the
universe quantum evolution. As the simplest example of the
suggested transition probability picture, we considered the
minisuperspace model for which classical and quantum evolution is
described by the harmonic vibrations  in conformal time
\bib{Lemos:1995qu}, \bib{Faraoni:1999qu}, \bib{gousheh}, \bib{Lapchinsky:1977vb}.
 The specific property of this minisuperspace model is that the
tomographic propagators for both classical and quantum universe
tomograms coincide. This fact provides some possibility to connect
observations related to today classical epoch of the universe  and
its purely initial quantum state. In the framework of the
suggested approach (and in the framework of the considered
oscillator model), the universe evolution can be studied using
specific properties of the tomographic propagator. If one
considers  tomograms and their evolution in classical mechanics
\bib{lecture}\bib{mendes2} the specific property of the linear systems (e.g.
oscillator model) is that the tomographic propagators in quantum
and classical domains are in one-to-one correspondence and are
given in the same carrier space, therefore we may say that the
difference of the quantum and classical evolution is solely in the
initial conditions, the quantum and classical pictures differ by
constraints which must satisfy quantum tomograms. They must
satisfy uncertainty relations. The choice of initial conditions
(initial tomogram of the universe) in correspondence with the
uncertainty relation provides a possibility to avoid the
singularity of the metric, which is unavoidable in the classical
picture. But the following evolution of the universe coded by the
tomographic propagator is the same (for the oscillator model).

Due to this result, one can extract from the present
observational classical data conclusions over the cosmological
evolution. Evolving backwards in time the present situation by
means of the ``true'' quantum or the classical propagators, we may
find discrepancies between the initial conditions at minus
infinity.

We are going to discuss this aspect in a future work.

\end{document}